\documentclass[conference,a4paper]{IEEEtran}
\usepackage{multicol}
\usepackage{breqn}
\usepackage{color}
\usepackage{soul}
\usepackage{amsmath}
\usepackage{graphics}
\usepackage{setspace}
\usepackage{cite}
\usepackage{latexsym}
\usepackage{float}
\usepackage{epsfig}
\usepackage{multirow}
\usepackage[table,xcdraw]{xcolor}
\usepackage{cite,cases,url}
\usepackage{amssymb}
\usepackage{graphicx}
\usepackage{epstopdf}
\usepackage{balance}
\usepackage{enumerate}
\usepackage{array}
\usepackage{algorithmic}
\usepackage{algorithm}
\usepackage{enumitem}

\usepackage[normalem]{ulem}
\useunder{\uline}{\ul}{}
\DeclareUnicodeCharacter{2212}{-}
\newcolumntype{L}{>{\centering\arraybackslash}m{5cm}}
\newcolumntype{K}{>{\centering\arraybackslash}m{6cm}}
\newcolumntype{P}{>{\centering\arraybackslash}m{2.3cm}}
\newcolumntype{M}{>{\raggedright\arraybackslash}m{2cm}}
\newcolumntype{N}{>{\raggedright\arraybackslash}m{2.5cm}}

\begin{document}

\pagenumbering{gobble}

\title{AI-Driven Demodulators for Nonlinear Receivers in Shared Spectrum with High-Power Blockers}

\author{
\IEEEauthorblockN{Hossein Mohammadi, Walaa AlQwider, Talha Faizur Rahman, and Vuk Marojevic
}%\\ %\vspace{0.2cm}
\normalsize\IEEEauthorblockA{Department of Electrical and Computer Engineering,  Mississippi State University,
Mississippi State, MS, USA}
$\{$hm1125$|$wq27$|$tfr42$|$vm602$\}$@msstate.edu%, vuk.marojevic@msstate.edu
\vspace{-3mm}
}

\maketitle

\begin{abstract}
Research has shown that communications systems and receivers suffer from high power adjacent channel signals, called blockers, that drive the radio frequency (RF) front end into nonlinear operation. Since simple systems, such as the Internet of Things (IoT), will coexist with sophisticated communications transceivers, radars and other spectrum consumers, these need to be protected employing a simple, yet adaptive solution to RF nonlinearity. This paper therefore proposes a flexible data driven approach that uses a simple artificial neural network (ANN) to aid in the removal of the third order intermodulation distortion (IMD) as part of the demodulation process. 
We introduce and numerically evaluate two artificial intelligence (AI)-enhanced receivers-ANN as the IMD canceler and ANN as the demodulator. Our results show that a simple ANN structure can significantly improve the bit error rate (BER) performance of nonlinear receivers with strong blockers and that the ANN architecture and configuration depends mainly on the RF front end characteristics, such as the third order intercept point (IP3). We therefore recommend that receivers have hardware tags and ways to monitor those over time so that the AI and software radio processing stack can be effectively customized and automatically updated to deal with changing operating conditions. 
\end{abstract}

\IEEEpeerreviewmaketitle

\section{Introduction}
\label{sec:intro}
Due to proliferation of smart applications, spectrum sharing has become a key enabler in order to accommodate massive number of heterogeneous devices running those applications. Spectrum sharing greatly enhances the efficient spectral utilization, allowing transmissions in compact adjacent spectrum bands \cite{zhang2018spectrum}. 
For example, the Citizens Broadband Radio Service (CBRS) band (3.55 - 3.7 GHz) in the United States has a three-tier spectrum access model where incumbent, priority access, and general authorized access users need to coexist in this new shared spectrum by following the spectrum access policy that are regulated by Federal Communications Commission \cite{munawwaCBRS}. 

In shared spectrum, the spectrum neighbors are not known a priori. Furthermore, spectrum agile devices usually have a wideband preselection filter, if any, i.e. a widely open radio frequency (RF) receiver front end. Software radios usually do no use a preselection %channel selection 
filter and filter the received signals in the digital domain. This means that the RF front end may potentially receive many undesired signals in bands that are adjacent to the signal of interest (SOI). % and if the sampling rate is not in line with the RF filter, aliasing may happen.

\begin{figure*}[ht]
\begin{center}
\includegraphics[totalheight=0.18\textheight]{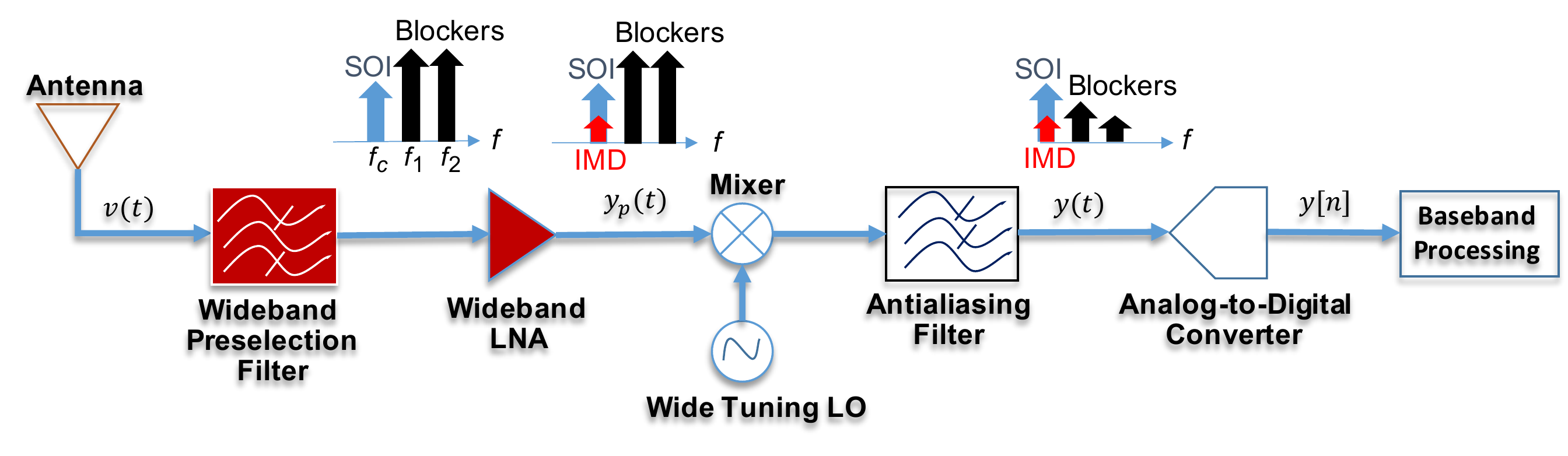}
\vspace{-2mm}
\caption{Receiver with nonlinear RF front end and the spectrum of the signal of interest (SOI), blockers, and intermodulation distortion (IMD) at the different stages of the receiver processing chain. The adjacent channel blockers, but not the IMD, %powers after the ADC 
can be eliminated by a conventional digital filter.}
\label{fig:fintermod}
\end{center}
\vspace{-5mm}
\end{figure*}

In this paper we consider the case of a wideband RF filter that encompasses multiple RF channels. The SOI occupies one channel. High-power adjacent channel signals can then act as blockers and may saturate the receiver or drive the RF front end into nonlinear operation. The worst case is saturating the analog-to-digital converter (ADC), which causes clipping and severe signal distortion. In the weak nonlinearity region, on the other hand, the low-noise amplifier (LNA) is driven in its nonlinear region causing signal compression. Two adjacent channel signals entering a nonlinear device, such as an LNA, cause intermodulation products. 
Typically, the most severe is the third order intermodulation, which is illustrated in Fig. \ref{fig:fintermod}. 
This is a well known problem and the two-tone test is employed to characterize and tag the quality of an RF component. The 3rd order intercept point (IP3), which has an input (IIP3) and output power (OIP3), is the metric that can be found in RF component data sheets.

The growing use of frequency agile software and cognitive radios, Internet of Things (IoT) devices, and RF sensors, as well as the ongoing trend of opening up traditionally licensed and protected spectrum for shared use motivates studying less controllable RF environments and their implications on communications performance. More precisely, since the shift is towards software solutions to compensate for hardware and channel impairments, nonlinearity effects that are externally triggered but affect radio receivers will need to be addressed in an efficient, yet adaptive way. 

Recent work has shown that even simple communications systems and receivers suffer from high power blockers \cite{dsouza2020symbol}. Since simple systems, such as the IoT, will coexist with sophisticated communications transceivers, radars and other spectrum consumers, these need to be protected employing a simple, yet adaptive solution. 
This paper therefore proposes a flexible data driven design that uses a shallow %designing and analyzing the use of an 
artificial neural network (ANN) to aid in the removal of the intermodulation distortion (IMD) as part of the demodulation process. 
We introduce and numerically evaluate two artificial intelligence (AI)-enhanced receivers, where in on case the ANN serves as a IMD canceler and in the other as the demodulator. Our results show that a simple ANN structure can significantly improve the bit error rate (BER) performance of nonlinear receivers with strong adjacent channel blockers and that the performance depends mainly on the receiver characteristics.

The rest of the paper is organized as follows: Section II formulates the problem and discusses the context and the related work. Section III introduces the system model based on well established theory and RF nonlinearity models. 
Section IV introduces our AI-enhanced receiver designs and configuration principles. 
Numerical results are provided and analyzed in Section V and conclusions are drawn in Section VI.

\section{Problem, Context, and Related Work}
\label{sec:problem}
\subsection{Problem}
%\textcolor{blue}{Brief context problem formulation}
In the context of spectrum sharing, where radio systems can operate in different channels and do not know their spectral neighbors a priori, high power signals in adjacent bands can cause nonlinear distortion to the desired signal at the receiver. This is especially true for low-cost receivers which have wide preselection filters and poor, i.e. highly nonlinear, RF front ends, such as typical Wi-Fi or IoT devices. We assume the case where such practical receivers operate in the weak nonlinear region, i.e. where there is no ADC clipping caused by the high-power blockers. 
%, which cause intermodulation products that fall in the band of interest and distort the signal. 
The problem is then to design and analyze a simple, yet adaptive %data driven
digital solution to demodulate the signal of interest in the presence of IMD.

\subsection{Context and Related Work}
The nonlinearity problem of components used in communications systems has been studied and different solutions have been proposed in RF and related communications contexts, including fiber optical systems. 
%RF nonlinearity has been %known and
%studied %for a while 
%in the past. % and known models and solutions exist.
The power amplifier (PA) is a major source of RF nonlinearity, where too high input signals drive it in its nonlinear operating region. PAs are expensive and inefficient and high linearity comes at a high cost \cite{9609963}. %PAs are needed for all communications systems. 
In order to avoid nonlinear operation, communications transceivers are carefully designed and tuned, and RF spectrum is carefully allocated, defining transmitter masks and considering adjacent frequency allocations. 
The tuning of communications systems often means that operators make sure to maintain sufficient input power backoff such that the input peaks are still withing the PA's linear operating region. The problem is exaggerated with advanced communications technology, specifically when employing multiple antenna systems %to increase data rates and higher communications reliability 
where each antenna element has a PA \cite{yaoICC18}. 
Digital predistortion is a common method to linearize the output of a PA \cite{agrawal2018rfpa}. 
%To achieve a higher data rates and communication reliability, the multiple-input multiple-output (MIMO) technology has been presented; however, radio frequency power amplifier (RFPA) is the main source of power consumption and nonlinearity impairment in the MIMO system. Reference \cite{agrawal2018rfpa}  employs DPD with disparate MIMO configurations achieves 6 times improvement in block error rate (BLER) and 2.77 times in throughput. 

%Spectrum bands are licensed for specific services, transmitter masks defined, and the adjacent band allocation is also considered. 
Receivers also consume spectrum in the sense that undesired signals that enter the RF processing chain, can distort the receiver and severely compromise its demodulation \cite{dsouza2020symbol} or sensing performance \cite{rebeiz2015spectrum}. Therefore, receivers need to get spectrum allocated and may need guard bands and limited accumulated power levels across the entire band that passes the preselection filter.

The premise of power backoff or careful spectrum regulation cannot be assumed for emerging wireless systems and spectrum regulation.
Specifically, 
cognitive radios can operate in different bands which means that they have to be able to receive signals in a wide range of frequencies. Software radios therefore have wide RF preselection filters, if any. 
Moreover, spectrum sharing among heterogeneous systems and services can cause high power neighboring signals that enter the receiver. 
Therefore, transceivers need to dynamically pre-process or post-process signals entering and exiting the nonlinear RF front end.

In \cite{yan2016digitally} an iterative IMD reconstruction has been proposed which outperforms a conventional method that uses least mean squares (LMS) filters to mitigate IMD. This algorithm can handle 5 dB higher adjacent channel interferers when compared with the LMS solution.

The main contribution of this work is that we analyze both high and low IMD effects stemming from high power blockers and moderate to highly nonlinear receivers. Moreover, as mentioned in \cite{yan2016digitally}, in contrast to LMS and iterative IMD in which the nonlinear parameter needs to be known, which requires characterizing the receiver, %assumption for practical situation
our approach estimates the IMD in real time during operation and removes it. %nonlinear part estimated and removed perfectly for different scenarios as we will see in simulation results. 
In addition, for algorithms such as LMS, determining the optimum step size is of %crucial
critical importance %owing 
to minimize the residual interference. %; however, this parameter is not available in the receiver side.

\section{System Model}
\label{sec:system}
Receiver nonlinearity has been studied and described by several researchers and the polynomial approximation model is widely used to describe it %non-linearity
\cite{razavi}%,McClan12}
:% The input-output relationship of the polynomial approximation is given by,
\begin{equation}\label{eqn:1}
    r(t) = \alpha_{1}x(t) + \alpha_{2}[x(t)]^{2} + \alpha_{3}[x(t)]^{3} + ..., %\alpha_{4}[x(t)]^{4} + ...,
\end{equation} 
% \end{align}
where $x(t)$ is the signal at the input of the nonlinear device and $r(t)$ is the output. Parameter $\alpha_{i}$ corresponds to the $i^{th}$ order gain (where $i$ = 1, 2, 3, ...) and $\alpha_{1}$ is called the linear gain. The third order nonlinearity usually causes the most distortion in the band of interest and the third order approximation is therefore often used \cite{dsouza2020symbol}.

For our system model, the desired signal, or SOI, is at frequency $f_c$ and the adjacent channel blocking signals are at $f_{1}$ and $f_{2}$ such that $2f_1-f_2=f_c$ (as in Fig. 1), or $2f_2-f_1=f_c$. The two adjacent channel blockers produce an intermodulation product at the frequency of the SOI when going through the nonlinear device, as shown in Fig.~1. %~\cite{razavi}.
Using (1), the received signal at $f_{c}$ can then be expressed as \cite{zou2009digital}
\begin{equation} \label{eqn:2}
r(t) = \alpha_{1}s(t) + \frac{3}{2} \alpha_{3} [b(t)]^{2} c(t)^{*} + n(t),
\end{equation}
%\hl{\textbf{\large Why the nonlinearity has defined like this: $[b(t)]^{2} c(t)^{*}$}}\\
%%\textcolor{red}{\begin{equation} \label{eqn:3}
%%Blocker =  \frac{3}{2} \alpha_{3} [b(t)]^{2} c(t)^{*} + n(t),
%%\end{equation}}
where $r(t)$ is the signal at the output of the nonlinear device, $s(t)$ is the transmitted signal, the SOI, $b(t)$ and $c(t)$ are the blocking signals, $n(t)$ is the additive white Gaussian noise (AWGN), and $\alpha_{1}$ and $\alpha_{3}$ are the linear and third order gains.

The third order intercept point, or $IP_3$, is a measure of RF component nonlinearity. Its input and output values can be obtained from data sheets of performing the two-tone test. The input $IP_3$ ($IIP_3$) % illustrated in Fig~\ref{fIIP3}. The input 
voltage and power can be calculated as \cite{rebeiz2015spectrum}, \cite{razavi}, \cite{zou2009digital}% corresponding to the $IP_{3}$ is given by
\begin{align}
A_{IIP_{3}} = \sqrt[]{\frac{4}{3} \bigg|\frac{\alpha_{1}}{\alpha_{3}}\bigg|}~~
\end{align}
and
\begin{align}
P_{IIP_{3}} = 20\log_{10}A_{IIP3} + 10 ~\text{[dBm]},
\end{align}
respectively.

\section{AI-Enhanced Practical Receiver}
\label{sec:contribution}
\begin{figure}
\centering
% \vspace{0}
% \hspace{-2pt}
\includegraphics[scale=0.26]{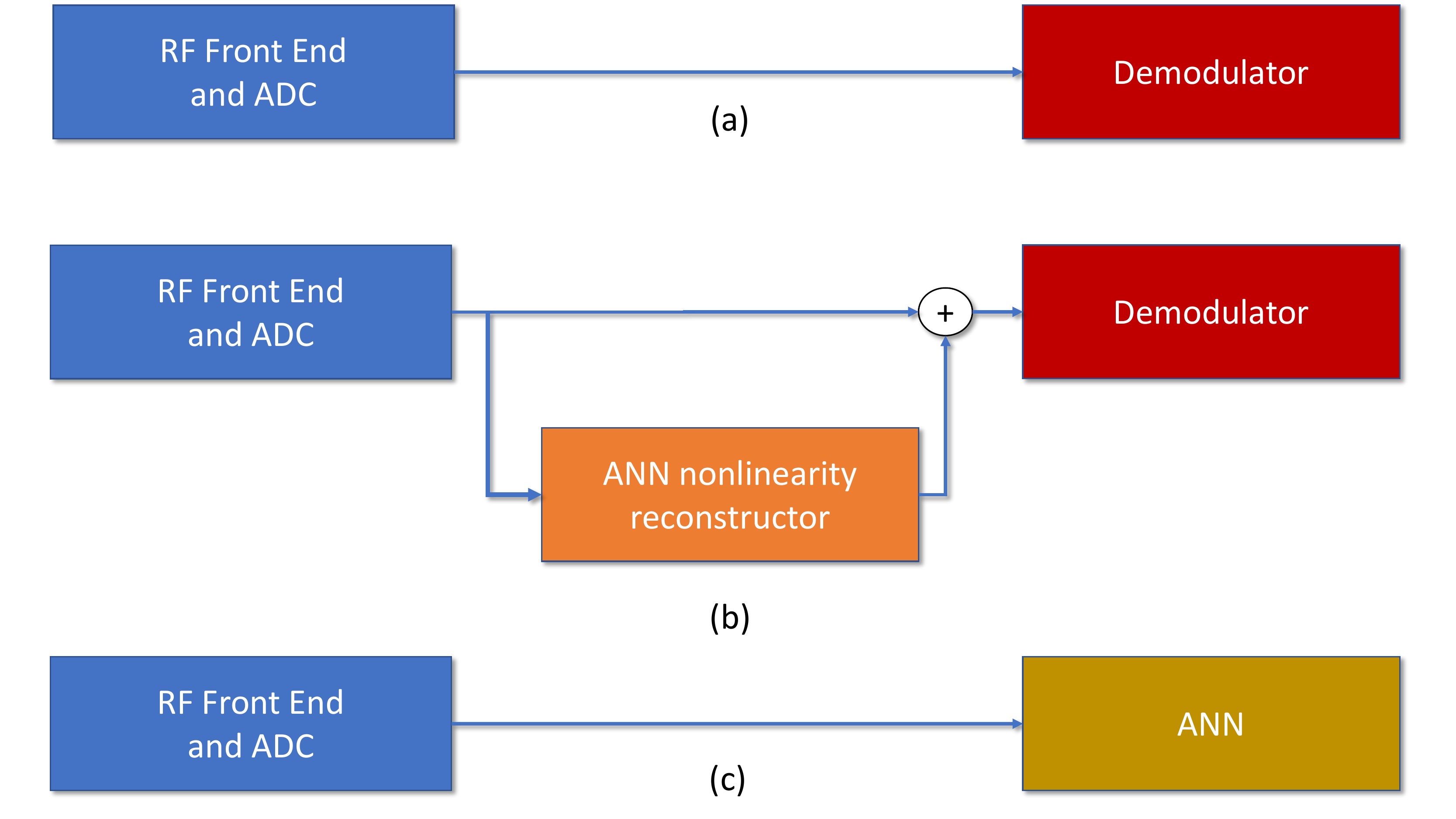}
% \hspace{-10cm}
\caption{Conventional receiver (a), ANN-based IMD canceling receiver (b) and ANN-based demodulator (c).}
\label{ANNModel}
 \vspace{-6pt}
\end{figure}

\subsection{AI Models}
%\textcolor{blue}{Short introduction to different AI models/algorithms that may apply here and what their strengths and weaknesses are with respect to the given problem.}\\

%Without the incorporation of AI having A entirely operative and efficient 5G network is not attainable. 
Existing 4G networks %, which are based on reactive conception, 
employ all-IP (Internet Protocol) broadband connectivity. %, this degrades the efficiency of using the spectrum. 
AI and its subcategories machine learning and deep learning have been evolving to the point that they will empower fifth-generation (5G) and Beyond 5G wireless networks %to be surmising and dynamic
\cite{osti_10287631, cayamcela2018artificial}. 
Moreover, in order to satisfy the growing %since the 
demand for wireless connectivity, % has grown exponentially over the last few decades, 
a new paradigm of wireless communications %, 6G systems, 
with the full support of AI is expected to be implemented between 2027 and 2030 \cite{chowdhury20206g}. 
For example, %In order to address one of the AI applications in the real world is optimizing the traditional data-driven approaches, V
vehicle to everything (V2X) communications %systems 
can benefit from AI to improve traditional schedulers and congestion control mechanisms and lower the current packet losses for % potential accidents and 
improved safety and comfort  \cite{tong2019artificial}.\\

Generally, AI techniques can be categorized as supervised and unsupervised learning. 
Machine learning and statistical logistic regression techniques, support vector machines (SVM) and artificial neural networks (ANN) are supervised learning techniques. %for the latter, models like 
K-means clustering and Q-Learning represent unsupervised learning. 
There is another approach named semi-supervised learning in which both labeled and unlabeled data exist for training the network.

\subsection{Proposed AI-Enhanced Receiver Designs}
%\textcolor{blue}{What model did we choose, why? How did we determine the design. How do we do the training.}
The proposed model in this paper is a supervised learning method in which a multilayer perceptron (MLP) ANN with only one hidden layer is designed to extract the information from the nonlinear part of the received signal by labeling the SOI as a target signal in the case where the desired signal is dominant. In contrast, for the case where the power of the interference is stronger than the SOI; by pausing the SOI transmission and just focusing on the nonlinear part of the received signal, the necessary information is extracted by the proposed ANN. 

Fig. \ref{ANNModel} shows three different receiver models considered in this paper. After the received signal passes the nonlinear front end and is digitized, the first receiver applies the conventional demodulator. % , the received signal after passes through the RF front end and the ADC and is applied directly to the demodulator. 
This is our baseline receiver. % for comparison in Section \ref{sec:simulations}. 
The first proposed AI-enhanced receiver uses an ANN nonlinearity reconstruction block (Fig. \ref{ANNModel}b). %, the received signal goes through the RF front end, ADC, and ANN nonlinearity reconstructor. 
Here the ANN tries to reconstruct the nonlinear part of (\ref{eqn:2}) to subtract it from the output of ADC before applying the conventional demodulation processes. 
The second proposed AI-receiver %third case is where 
designs the ANN to take a distorted input signal and demodulate it. %when the SOI and the blocker powers are moderate. 
The goal of this %third
receiver is to evaluate whether the ANN can directly extract and demodulate the SOI from the distorted received signal. 
Our %basic 
ANN structure is
%Below we clarify the inputs and the target signals:
\begin{equation}
    \mathrm{net = ANN\{Input,Target,Learning Algorithm\}.}
\end{equation}

{For quadrature modulated symbols, we have two options, work with two ANNs, one for the in-phase (I) and the other for the quadrature (Q) component, or augment the basic ANN structure as follows:} %since there is a real and an imaginary part, % of signals should be separated, 
%we augment these elements together as follows:
\begin{equation}
\begin{aligned}
    net = ANN\{[\Re (In),\Im (In)],
    & [\Re(T),\Im(T)],LA\},
\end{aligned}
\end{equation}
where $\Re(.)$ and $\Im(.)$ stand for the real and imaginary parts, that is, the I and Q components. $In$, $T$, and $LA$ represent the input and target signals, and the learning algorithm.

%%\textcolor{red}{\begin{equation}
 %% net = ANN\{[real(r(t)),imag(r(t))],
 %%   & [real(Blocker),imag(Blocker)]\\  & ,BNN\}.
%%\end{equation}}

{For the problem considered in this paper our basic ANN has one neuron in the input layer and one neuron in the output layer with one hidden layer that has four neurons. The activation function is $tansig$ for the hidden layer and $purelin$ for the output layer.
This %basic 
ANN is used for implementing the ANN canceler (Fig. \ref{ANNModel}b) and ANN demodulator (Fig. \ref{ANNModel}c) for a BPSK receiver. The weights are trained as discussed in the next paragraph. 

For the QPSK case, since we have both I and Q components, we use 2 input and 2 output neurons for the ANN canceler, with one hidden layer with four neurons and activation %, and the activation
functions as before. % is $tansig$ for the hidden layer and $purelin$ for the output layer. 
For the ANN demodulator, we apply one ANN in the I and one in the Q path where each has %are %and both are
%of the same 
the same structure as the basic ANN designed for BPSK demodulation.}

%As mentioned before in Fig. \ref{ANNModel}b, 
%\textcolor{blue}{For the ANN canceler (Fig. \ref{ANNModel}c), the input to the ANN is (\ref{eqn:2}) without the desired signal (SOI absent, null pilots). The output target is be the nonlinear part. The noise cannot be removed and will always be part of the system. 
%For the ANN demodulator, (Fig. \ref{ANNModel}c), the input is be the received signal with SOI pilots as the labeled data and the target is the SOI.}

\subsection{Control Signals and Training}
\textcolor{black}{As explained in \cite{bishop2006pattern}, the MLP tries to minimize the mean square error and fit the curve to the labeled data points. Therefore, for ANN canceler (Fig. \ref{ANNModel}b), the input to the ANN is %(\ref{eqn:2}), %and in case of high blocker powers, 
%the input is 
the nonlinear component of (\ref{eqn:2}) plus noise, where the learning target is the nonlinear component. In this case, the SOI is absent; that is, the transmitter is turned off. In other words, the SOI pilots are zero. 
Note that the ANN is not interacting with a pure signal, the nonlinear component of (\ref{eqn:2}) in this case. This makes the training process reliable and practical. }
%The AWGN cannot be removed and will always be part of the system. Because of its randomness, it cannot be the target for learning. 
%We train the network with $SNR = 5 dB$ to not overfit the ANN.

\textcolor{black}{For the ANN demodulator (Fig. \ref{ANNModel}c), the SOI transmits regular data known to the receiver during the training phase where the target is the SOI. That is, pilots are established as part of the control signaling between transmitter and receiver and are used as the labeled data for training.
The trained ANN demodulator %thus
then tries to extract the transmitted bits from the SOI which it receives together with the in-band nonlinear distortion and noise.
}

\subsection{Learning Algorithm}
The Bayesian Regularization (BR) algorithm is chosen for training for a few reasons. First, as discussed in  \cite{bishop2006pattern}, the two hyperparameters in a Bayesian Neural Network (BNN) result in utilizing the effective number of weights while training the network. 
%\textcolor{red}{The reason of choosing this algorithm stems from the fact that since we have no idea about the system complexity and modulation type for some cases, it is not necessary to use all of the weights for learning; therefore, our ANN may not be a fully connected which makes the system more complex and may cause errors during the training process.}
The error is a function of the weights and the hyperparameters  %parameters 
$\alpha$ and $\beta$:
\begin{equation}
    E(w_{MAP}) = \frac{\beta}{2} \sum_{n=1}^{N} \{y(x_n,w_{MAP})-t_n\}^2 + \frac{\alpha}{2}w_{MAP}^T w_{MAP}.
    \label{eq7:BN}
\end{equation}
%To elaborate more on (\ref{eq7:BN}) 

In order to minimize the error function we need to %optimize determine
tune the values of parameters $\alpha$ and $\beta$. Vector $w_{MAP}$ indicates the posterior distribution of the network weights. 
Since we are obtaining the initial values for the weights from a normal distribution, some of these weights might be useful during the training at first, but may add error %at the end 
to the system. Therefore, a trained ANN may end up not being fully connected. For more information please refer to \cite{bishop2006pattern}

%\balance
\section{Numerical Analysis}
\label{sec:simulations}
%\textcolor{blue}{Organize it such to allow a systematic analysis. Sample subsections below.}

The simulator uses complex baseband representation of the transmitted signal. The received signal is the attenuated signal plus the IMD as a result of co-channel blockers and receiver nonlinearity plus AWGN according to (2). The total noise power is set as $N$ = −114 dBW for a bandwidth of $BW = 1$ MHz to account for noise floor elevation due to co-channel interference and receiver noise figure. %\{\textcolor{red}{ A unit linear gain is assumed.(this is the value of $\alpha_1$) I cannot understand this}\} 

%Each term is represented by voltages and the symbol rate is constant. 
We assume that the transmitted and blocker signals all %are M-PSK modulated and 
experience an AWGN channel. $P$, $B$, and $C$ are the desired and the two blocker signal powers, respectively, at the receiver. % of interest. 
%We capture them in relation to the noise floor. 
The bits of the desired signal and the adjacent channel blockers are produced by independent random sources.

We consider the three receiver implementations from Section \ref{sec:contribution} and shown in Fig. 2: the conventional demodulator, the ANN-based IMD canceler, and the ANN demodulator. 
We also consider %include the results for 
the conventional receiver without nonlinearity in an AWGN channel. %without blocker. 
The power level of each blocker is set to a high value of 70 dB above the noise floor to evaluate the case of high IMD. We first choose $IIP_3$ = -10 and -20 dBm to model two types of nonlinear receivers. We then evaluate the performance of the different receivers as a function of the nonlinearity figure between an IIP3 of -30 and +20 dBm, providing a range of practical nonlinear receivers. % from poor to excellent. % to evaluate the performance for different receiver nonlinearity regions. 
The SOI and blockers are both assumed to carry data and employ binary and quadrature phase shift keying (BPSK and QPSK) modulation. 

\begin{figure}
\begin{center}
\includegraphics[width=1.04\linewidth]{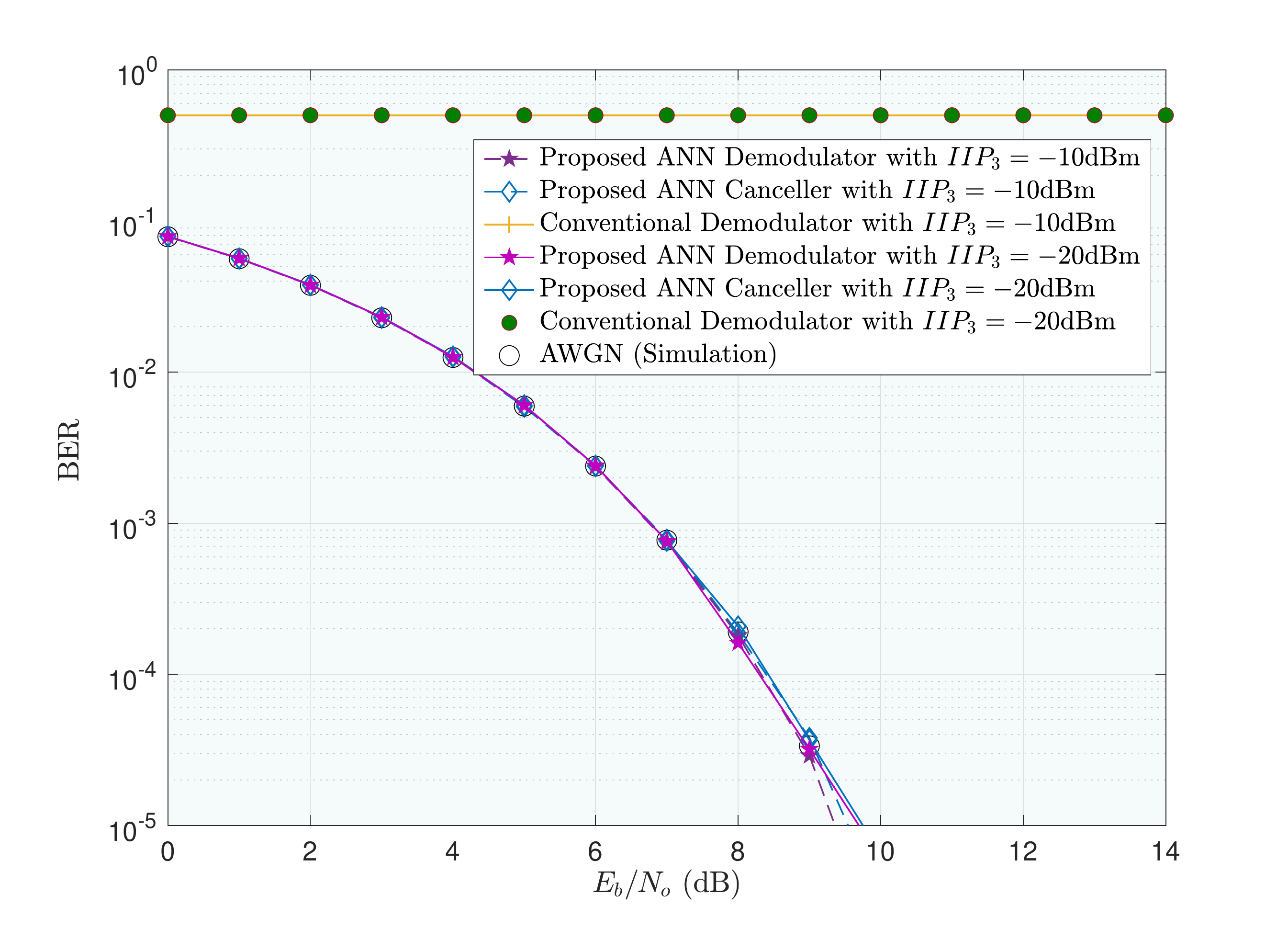}
\vspace{-5mm}
\caption{BER of a BPSK modulated signal with adjacent channel blockers and receiver nonlinearity. %All signals---desired and blockers---
The blockers are BPSK modulated signals and are received at %with the blocker powers being 
70 dB above the noise floor. %We assume two IIP3 values, representing poor receivers. 
The AWGN curve represents the conventional demodulator for an ideal (linear) receiver.}
\label{fig:result1}
\vspace{-5mm}
\end{center}
\end{figure}

\begin{figure}

\begin{center}
\includegraphics[width=1.04\linewidth]{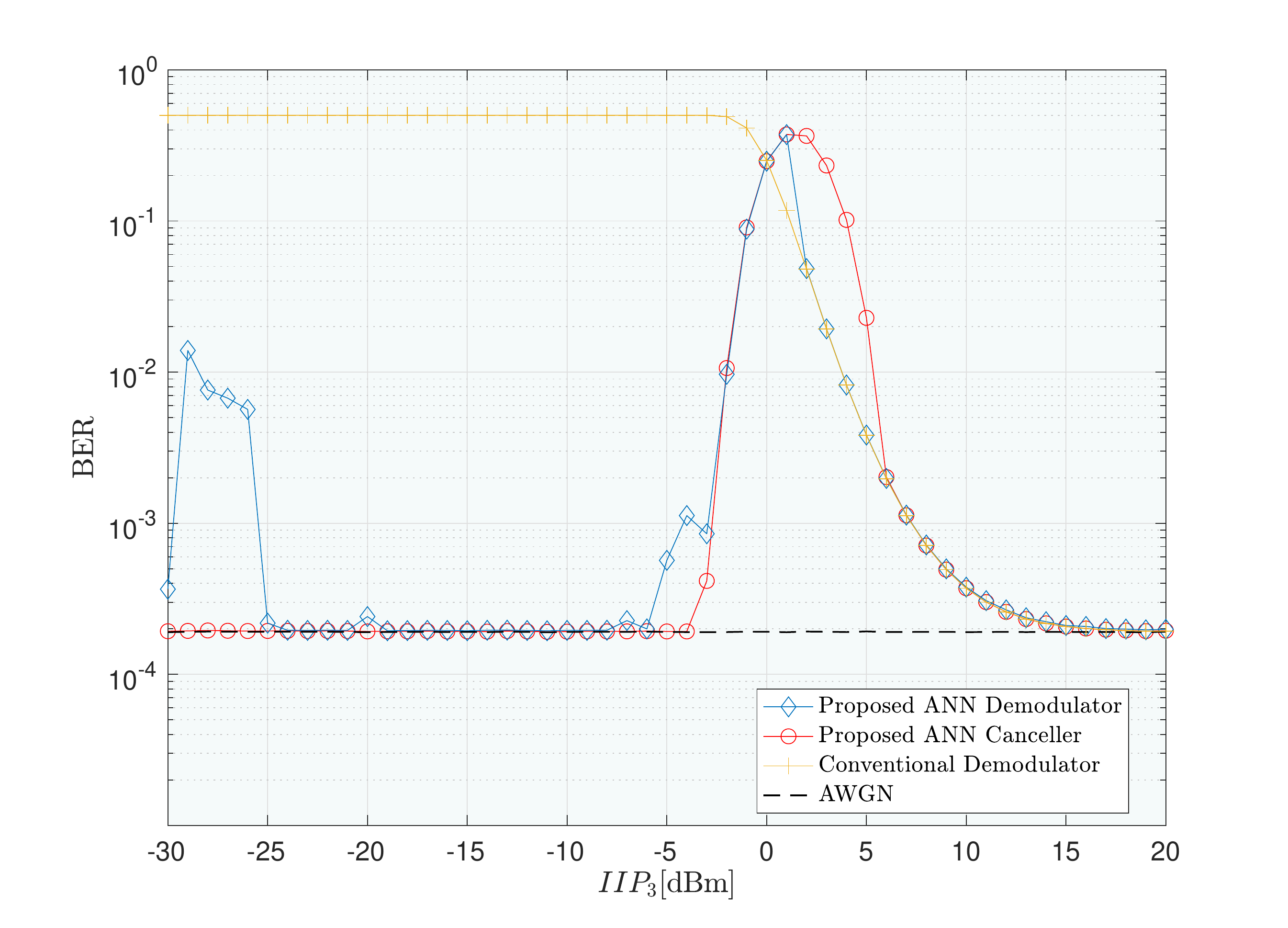}
\vspace{-5mm}
\caption{BER for a BPSK signal for the three nonlinear receivers of Fig. 2 as a function of $IIP_3$ for $E_b/N_0$ = 8 dB. The ideal receiver in AWGN channel is shown for reference.}
\label{fig:result4}
\vspace{-4mm}
\end{center}
\end{figure}

\subsection{BPSK SOI and Blockers for Different $IIP_3$ Values}
 
Fig. \ref{fig:result1} plots the BER of BPSK communications over $E_b/N_0$ for the different receivers and two IIP3 values with blocker power levels of 70 dB above the noise floor. 
The conventional demodulator curve shows how the nonlinearity affects and considerably degrades the BER performance at high blocker power level, even for a BPSK modulation scheme. %, which is robust. 
Also, the figure shows that in general the two proposed ANN models improve the system performance and completely eliminate the effect of the nonlinearity for both IIP3 values, -10 and -20 dBm, which model highly nonlinear receivers. 
%Independent of the IIP3 value, 
%The performance of 
Both AI-enhanced receivers %methods 
are able to %essentially
mitigate the IMD and closely match the AWGN lower bound. 
%better than the case when IIP3 = -10 which concludes that the ANN is good when the quality of the receiver is poor based on the nonlinearity. More specifically 
The ANN canceler %model
slightly outperforms the ANN demodulator at high SNR values. % for both IIP3 values. %\{\textcolor{red}{I think it's better to zoom in this part of figure}\}

Fig. \ref{fig:result4} compares the BER performances %of the BPSK communications system %modulation 
of the three receivers of Fig. 2 for BPSK and an $E_b/N_0$ of 8 dB as a function of %the
$IIP_3$ with B = C = $N_0$ + 70 dB blocker powers. It shows that the conventional demodulator completely fails to extract and demodulate the data from the distorted signal for moderate and highly nonlinear receivers ($IIP_3$ below 0 %-5 to -30 
dBm), whereas the ANN canceler %for the same region of nonlinerarity 
is able to completely eliminate the severe IMD that otherwise makes the demodulation impossible. The ANN demodulator shows excellent %good
performance, %on the mentioned nonlinearity region 
but with some instability for highly nonlinear devices. % high nonlinear region. In the 

The performance gains of the AI-enhanced receivers stem from the fact that high IMD %nonlinearity 
causes the ANN to interact with an almost pure input signal; therefore, by estimating an appropriate function it can fit the curve to the received signal. For low IMD, the AI-enhanced receivers converge to the AWGN benchmark. At intermediate IMD, the proposed cancellation/demodulation techniques behave similar %(slightly worse) 
to the conventional receiver. %, which does not combat nonlinear distortion. 
%%As a corollary, %For more elaboration on this, 
%%assume that we have a noisy sine wave. If the variance of the noise added to the signal is high, predicting and fitting a curve to the original signal is almost impossible.\\
% Vuk: I did not understand the corollary and what it explains here. 
 
\subsection{QPSK SOI and Blockers for Different IIP3 Values}
\begin{figure}
\begin{center}
\includegraphics[width=1.04\linewidth]{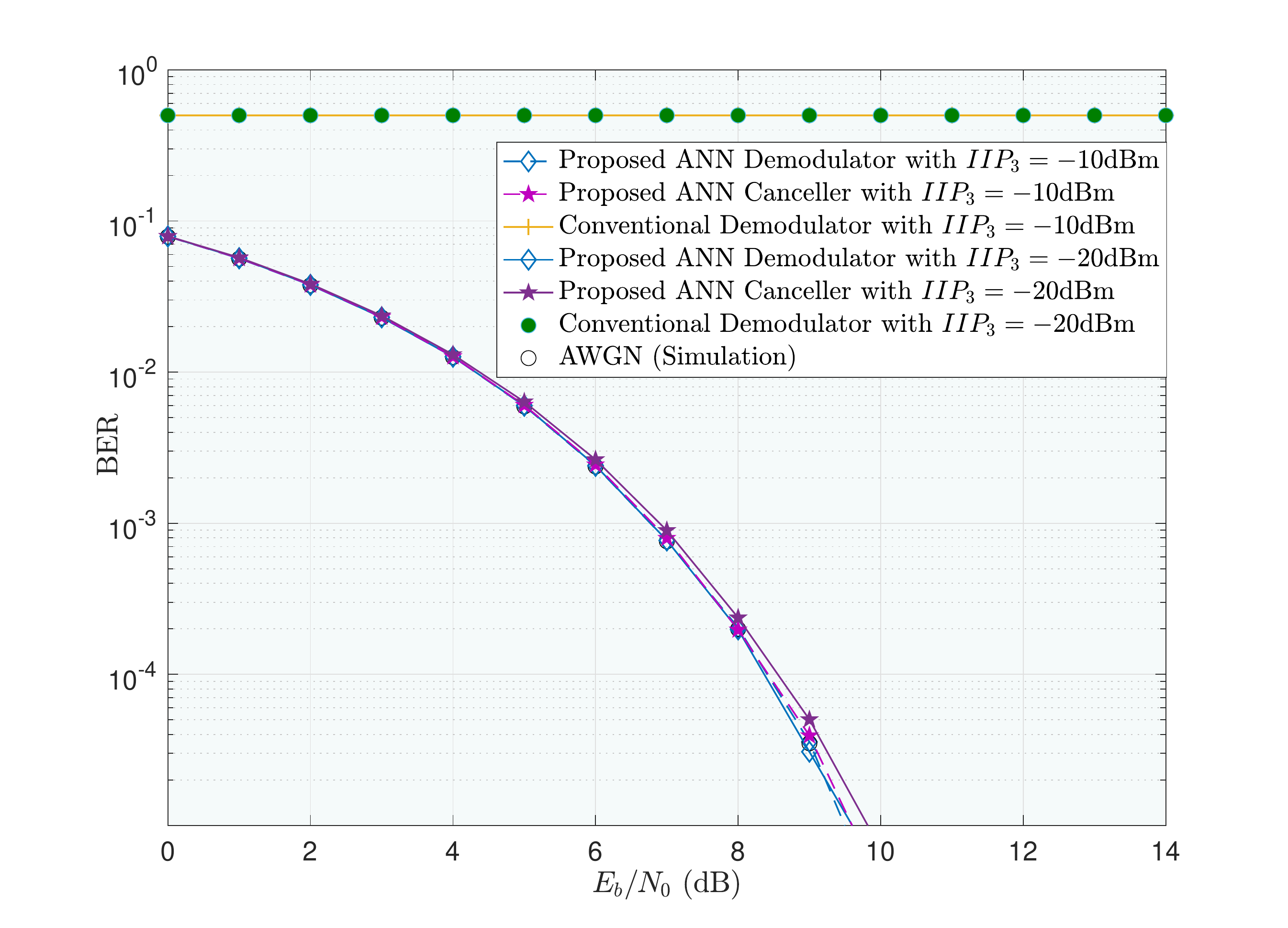}
\vspace{-5mm}
\caption{BER of a QPSK modulated signal with adjacent channel blockers and receiver nonlinearity. %All signals---desired and blockers---
The blockers are QPSK modulated signals and are received at %are  modulated with the blocker powers being 
70 dB above the noise floor. %We assume two IIP3 values, representing poor receivers. 
The AWGN curve represents the conventional demodulator for an ideal (linear) receiver.}
\label{fig:result2}
\vspace{-5mm}
\end{center}
\end{figure}

Fig. \ref{fig:result2} shows the simulation results for QPSK modulated SOI and blockers. We notice that both ANN solutions are able to effectively remove the effect of nonlinearity and match the BER curve of the AWGN system. 
The increase of dimension from BPSK to QPSK adds a dimension to the ANN input, but %because of our ANN designs, 
the gains of the proposed AI-enhanced receivers do not suffer and their performances still closely match the AWGN lower bound   %Even with the increasing in the modulation order which should add more error probability to the system 
%the ANN-enhanced receivers %still 
%perform excellent in mitigating the IMD 
for both receiver types (IIP3 = -10 and -20 dBm).   % modulator is used for both the SOI and blockers. 
%Reference \cite{dsouza2020symbol} proved that the BER performance gets more affected by the nonlinearity if the SOI modulation order increases and this is also observed from this figure when compared to the previous. % \ref{fig:result2} if we compare it to figure \ref{fig:result1}. 
%The conventional receiver is therefore at the error floor then also reaching the 
%The gains of the ANN-enhanced receiver %diminish solutions over the conventional receiver is 
%are diminished compared to the BPSK case, but are still significantly better than the error floor of the conventional demodulator. % especially for the ANN canceler model. 
%Although the ANN canceler does not eliminate the total effect of the nonlinearity as it does for the BPSK case, it provides reasonable %still %does a good job in
%improves the 
%performance with a 2 loss compared to the ideal receiver. 
%There is hence a 4 dB SNR loss for receiver front ends characterized by an $IIP_3$ of -20 dBm compared to an $IIP_3$ of -10 dBm. %in comparison with its counterpart IIP_3 = -10dBm refers to the case that a fixed ANN structure has been used in all cases; 1 hidden layer with 4 neurons.
%of the system BER for both values of IIP_3. 
%The ANN demodulator model, although better than the conventional demodulator, shows a poor performance in the case of QPSK. 
%In addition, 

\begin{figure}
\begin{center}
\includegraphics[width=1.04\linewidth]{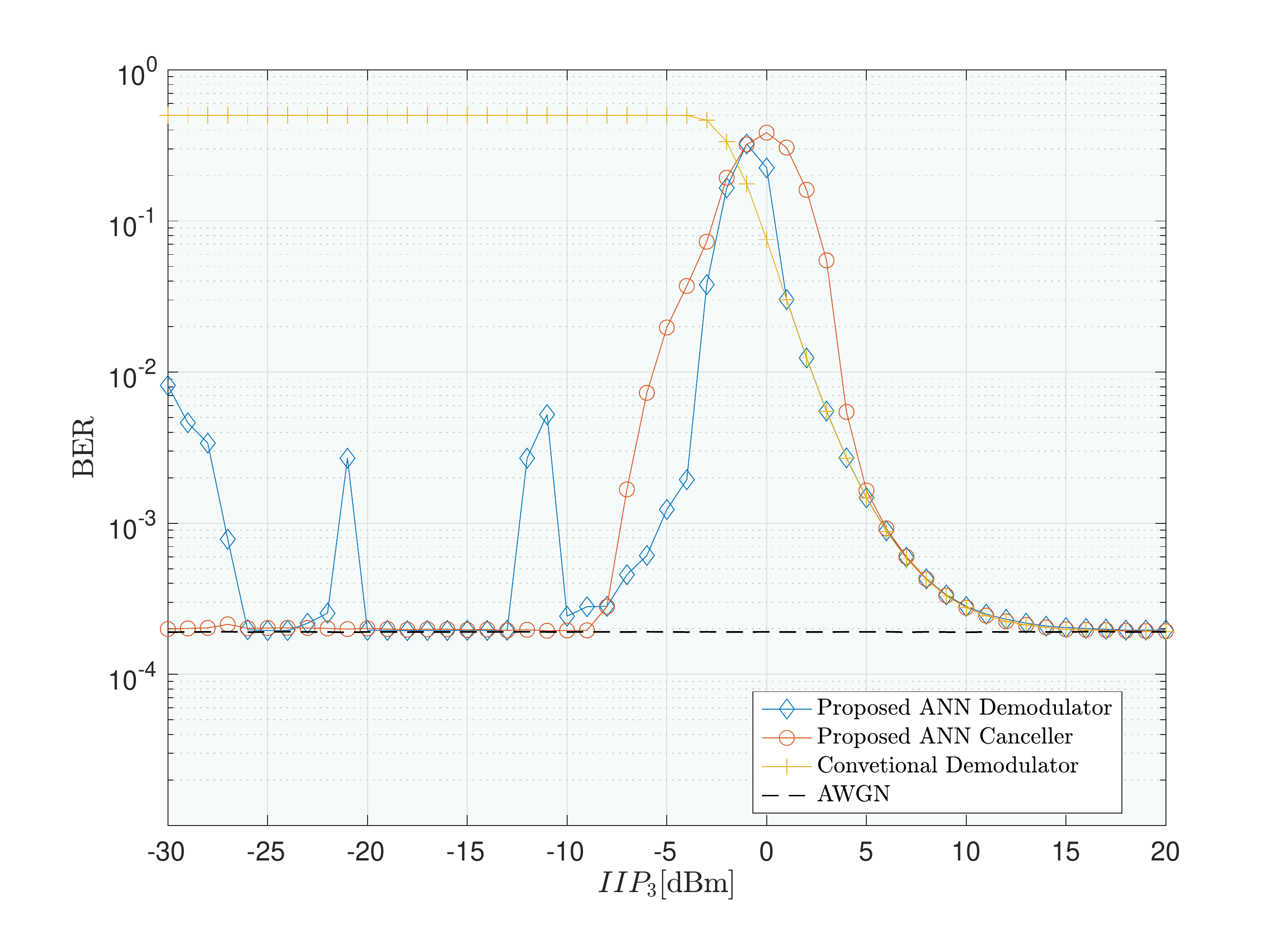}
\vspace{-8mm}
\caption{BER for a QPSK signal for the three nonlinear receivers of Fig. 2 as a function of $IIP_3$ for $E_b/N_0$ = 8 dB. The ideal receiver in AWGN channel is shown for reference.}
\label{fig:result3}
\vspace{-5mm}
\end{center}
\end{figure}

%\textcolor{red}{\hl{Fix this paragraph based on the updated results:}}
Fig. \ref{fig:result3} shows the performances of the three receivers of Fig. 2 for an $E_b/N_0$ of 8 dB as a function of $IIP_3$ with B = C = $N_0$ + 70 dB blocker powers. These results confirm that for moderate and highly nonlinear receivers, the ANN canceler outperforms the other structures and is able to {completely eliminate the severe IMD that otherwise makes the demodulation impossible}. 
So does the ANN demodulator, but we observed outliers which may be due to the initialization of weights during training. Out of 100 ANN instances only few show anomalies and orders of magnitude higher BER which pull the result up, as observed by the three peaks in the ANN demodulator curve at IIP values around -30, -21 and -11 dBm.

The results for $IIP_3$ between -5 and +5 dBm need to be further analyzed. We call this the medium nonlinear region. What we observe here is that the ANN demodulator performs better in aiding in the demodulation than the ANN canceler. The IMD is less severe here and the ANN canceler has trouble approximating it for removal. The ANN demodulator, on the other hand, which tries to approximate the SOI, is able to handle it better and closely approaches the conventional demodulator curve. % around 0 dBm IIP3, which the ANN canceler approximates at +5 dBm. 
Where the IMD is low, for receivers with IIP3 of +5 dBm or more, the ANN processing is not needed, but it does not disturb the demodulation process.

The training of the MLP ANN is performed using matrices in MATLAB. This simplifies the derivation of the computational complexity as function of the number of layers and the number of training samples. With $t$ training samples, the order of complexity of operation while propagating from one layer with $m$ nodes to the next layer with $n$ nodes is, $\mathcal{O}\left ( t \ast m \ast n \right )$. Hence, with one input layer with $m$ nodes, one hidden layer with $n$ nodes and one output layer with $k$ nodes, the complexity of proposed ANN receiver for one epoch is, $\mathcal{O}\left ( t \ast m \ast n \ast k \right )$. Since $I$ epochs are required to train the network, the overall complexity becomes, $\mathcal{O}\left ( t \ast I \ast m \ast n \ast k \right )$, which corresponds to the number of matrix multiplications.

%The complexity of the proposed method depends on the ANN structure. This complexity can be calculated as a function of the number of layers and neurons in each layer. The time complexity of one iteration is $O(D)$, where $D$ is found to be $D\triangleq \sum_{l=1}^{L}n_ln_{l+1}$ \cite{comlex}, $N$ is the number of layers, and $n_l$ is the number of neurons in layer $l$. For the training phase, the complexity is $O(t*I*D)$, where $t$ is the number of training samples and $I$ is the number of iterations to train the network.}

In conclusion, our results show that the AI-aided demodulation process is robust. It enables demodulation with severe IMD %nonlinearity (without ADC clipping, which we have not analyzed here) 
which is otherwise impossible. For medium IMD levels, %ity 
the two AI-enhanced receivers %it is 
perform similar %close
to the regular receiver. %performance 
For excellent receivers or low blocker powers, where the proposed ANN processing is not needed, it provides the expected demodulation performance. 
The region of medium IMD needs to be further analyzed. The ANN structure and training can be further optimized for such receivers and operating scenarios. 
%To address a solution to this degradation in performance, first we need to consider this fact transferring from BPSK which is one dimensional modulation to QPSK, means we are adding dimension to the ANN input; therefore, without a doubt the performance decreases.\\
%\newline

\subsection{Discussion}
In this paper we have designed an ANN with a fixed structure, which is characterized by one hidden layer with four neurons. 
One compelling solution to improve the design is to consider %this problem which could be an intriguing topic for future research is using 
an adaptive ANN (AANN). 
The AANN allows defining different ANN structures with different training algorithms. It can %thus
adapt itself to observe the defined criteria. 
For instance, as Fig. \ref{fig:result2} shows, %\{\textcolor{red}{again figures 4 and 5 dsplaced}\} 
the proposed ANN is trained very well by reaching one of the defined parameters; gradient, performance or epoch, then based on the trained network the system is used to aid in the demodulation and establishes a reasonable BER performance. 
On the other hand, the AANN can define a different constraint, such as having an SNR loss of less than 2 dB. Then the proposed ANN begins to switch between different structures and algorithms, e.g. considering the Stochastic Gradient Descent Momentum (SGDM), with the aim of meeting the aforementioned constraint. Although this approach applies a very high complexity to the system, it ensures that the results are the optimum among a variety of structures.
%\{\textcolor{red}{BPSK is shown in Fig.~\ref{fig:result4}}, I think it is out of topic\}

%\textcolor{red}{One paragraph: Write about the results and what is happening and why or what might be the problem.}

%I forgot about it :( sorry

%\textcolor{red}{One or two paragraphs: These results are based on a fixed ANN structure, fixed learning anf fixed activation functions. An emerging research is to use adaptive AI, which can find the optimal structure for a challenging problem...<explain the options>}

\balance
\section{Conclusions}
\label{sec:conclusions}
We have proposed %In this paper 
two AI-enhanced receivers %models were proposed 
to mitigate the effects of the RF nonlinearity of practical %and poor quality 
systems. Two communications systems %different PSK modulation orders 
have been examined using the proposed methods and the results evaluated against the ideal system and the conventional receiver. 
The simulation results first show that the conventional receiver cannot demodulate the data with adjacent high-power blockers causing moderate or high IMD due to receiver nonlinearity. A significant performance improvement of the two proposed ANNs---IMD canceler and demodulator---over the conventional receiver is numerically shown when the blocker power levels are high. % for BPSK modulated signals. %ion is used for both SOI and blocker signals. 
%For QPSK modulated signals, only the ANN canceler model significantly improves the system performance by mitigating the IMD. \{\textcolor{red}{we showed that demodulator can still work if we change the ANN structure}\} 
We conclude that the use of the proposed ANN nonlinearity canceler, which focuses on estimating and removing the IMD caused by high-power blockers, as well as the ANN demodulator can significantly improve the BER performance. The ANN canceler outperforms the ANN demodulator in terms of stability for highly nonlineary receivers and allows communications even in severely nonlinear systems and harsh signaling environments. The ANN demodulator, on the other hand, outperforms the ANN canceler for medium %severe
IMD. %nonlinearity in the receiver RF chain. 
Both match the conventional receiver performance when the IMD is low or %when it is 
negligible. 
%when the blocker power and the SOI and blockers modulation order is high. 

This research has shown that employing data-driven processing with ANNs, which can inherently model nonlinear behavior, can effectively help mitigating hardware typical RF impairments. We therefore recommend considering AI controllers for driving future transceivers, both for simple and low-cost IoT devices as well as for sophisticated broadband wireless networks. %Future
Research needs to analyze and experimentally verify the applicability of AI and devise suitable structure and their stability for controlling %or implementing 
modern broadband waveforms and practical communications systems.

\section*{Acknowledgment}
This work was supported in part by the National Science Foundation grants  %number
CNS-1564148, % and
ECCS-2030291 and CNS-2120442. 

\bibliographystyle{IEEEtran}
\bibliography{main}

\end{document}